\documentclass[sigconf, authorversion]{acmart}
 
\acmConference[ESEM]{}{2020}{ACM}

\usepackage{booktabs} 
\usepackage{graphicx}

\usepackage{xspace}
\usepackage{graphicx}
\usepackage{ifthen}
\usepackage[normalem]{ulem} 
\usepackage{xcolor,colortbl}
\usepackage{hyperref}

\newboolean{showedits}
\setboolean{showedits}{true} 
\ifthenelse{\boolean{showedits}}
{
	\newcommand{\del}[1]{\textcolor{red}{\sout{#1}}} 
	\newcommand{\nbe}[3]{
		{\colorbox{#3}{\bfseries\sffamily\scriptsize\textcolor{white}{#1}}}
		{\textcolor{#3}{\sf\small$\blacktriangleright$\textit{#2}$\blacktriangleleft$}}}
}{
	\newcommand{\del}[1]{} 
	
	\newcommand{\nbe}[3]{}
}

\newboolean{showcomments}
\setboolean{showcomments}{true} 
\newcommand{\id}[1]{$-$Id: scgPaper.tex 32478 2010-04-29 09:11:32Z oscar $-$}

\ifthenelse{\boolean{showcomments}}
 {
 	\newcommand{\nbc}[3]{
 		{\colorbox{#3}{\bfseries\sffamily\scriptsize\textcolor{white}{#1}}}
		{\textcolor{#3}{\sf\small$\blacktriangleright$\textit{#2}$\blacktriangleleft$}}}
	
 }{
 	\newcommand{\nbc}[3]{}
 	
 }

\usepackage[most]{tcolorbox}
\ifthenelse{\boolean{showedits}}
{
  \newtcolorbox{inserted}{%
       title=Inserted text:,
       colframe=blue,colback=blue!5!white,
       breakable,
       leftrule=0mm, 
       bottomrule=0mm,
       rightrule=0mm,
       toprule=0mm,
       arc=0mm, outer arc=0mm,
       oversize
  }
  \newtcolorbox{deleted}{%
       title=Deleted text:,
       colframe=red,colback=red!5!white,
       breakable,
       leftrule=0mm, 
       bottomrule=0mm,
       rightrule=0mm,
       toprule=0mm,
       arc=0mm, outer arc=0mm,
       oversize
  }
  \newtcolorbox{refactored}{%
       title=Rewritten text:,
       colframe=blue,colback=red!5!white,
       breakable,
       leftrule=0mm, 
       bottomrule=0mm,
       rightrule=0mm,
       toprule=0mm,
       arc=0mm, outer arc=0mm,
       oversize
  }
}{

}

\newcommand{\commented}[1]{}

\newcommand{\eg}{\emph{e.g.,}\xspace}
\newcommand{\ie}{\emph{i.e.,}\xspace}
\newcommand{\etal}{\emph{et al.}\xspace}

\newcommand{\code}[1]{\texttt{#1}}
\newcommand{\MD}{\code{MessageDigest}\xspace}

\usepackage{listings}
\usepackage{color, colortbl}
\usepackage{multirow}
\usepackage{adjustbox}

\lstdefinelanguage{Java}{
  tabsize=4
}[keywords,comments,strings]

\definecolor{source}{gray}{0.95}
\definecolor{highlight}{gray}{0.9}

\lstnewenvironment{codesnippet}{%
	\lstset{%
		frame=single,
		framerule=0pt,
		mathescape=false
	}
}{}

\newcommand{\boxit}[1]{\vspace{0.3cm}
\noindent
\fbox{
\begin{minipage}{25em}
\emph{#1}
\end{minipage}
}
}

\newcommand{\CG}{CogniCrypt\xspace}
\newcommand{\GH}{GitHub\xspace}

\newcommand{\CM}{CryptoMine\xspace}

\newcommand{\vs}{vs.\xspace}

\makeatletter                  
\def\mdseries@tt{m}      
\makeatother                   

\begin{document}
\title{Java Cryptography Uses in the Wild}



%

\author{Mohammadreza Hazhirpasand}
\affiliation{%
  \institution{University of Bern}
  \city{Bern}
  \state{Switzerland}
}

\author{Mohammad Ghafari}
\affiliation{%
  \institution{University of Bern}
  \city{Bern}
  \state{Switzerland}
}

\author{Oscar Nierstrasz}
\affiliation{%
  \institution{University of Bern}
  \city{Bern}
  \state{Switzerland}
}

%
%

\begin{abstract}
[Background] Previous research has shown that developers commonly misuse cryptography APIs. [Aim] We have conducted an exploratory study to find out how crypto APIs are used in open-source Java projects, what types of misuses exist, and why developers make such mistakes. [Method] We used a static analysis tool to analyze hundreds of open-source Java projects that rely on Java Cryptography Architecture, and manually inspected half of the analysis results to assess the tool results. We also contacted the maintainers of these projects by creating an issue on the GitHub repository of each project, and discussed the misuses with developers. [Results] We learned that 85\% of Cryptography APIs are misused, however, not every misuse has severe consequences. Developer feedback showed that security caveats in the documentation of crypto APIs are rare,  developers may overlook misuses that originate in third-party code, and the context where a Crypto API is used should be taken into account. [Conclusion] We conclude that using Crypto APIs is still problematic for developers but blindly blaming them for such misuses may lead to erroneous conclusions.

\end{abstract}

\keywords{Java cryptography, security, empirical study}

\maketitle

\section{Introduction}
\label{sec:intro}
Cryptography is the primary mechanism for data protection in all privacy and security-sensitive services available in the digital world.
Nevertheless, cryptography APIs (or ``crypto APIs'') are not easy to use for developers.
A recent study reported that 72\% of Java projects on \GH suffer from at least one crypto misuse such as weak algorithm selection, short encryption key size, or obsolete function calls~\cite{hazhirpasand2019impact}.

The lack of proper security-related hints in the common information sources for developers aggravates this issue.
Oracle's official online documentation of the Java Cryptography Architecture (JCA) is mostly limited to the explanation of each API and its parameters, and it rarely explains caveats about secure configuration of the APIs.
%
For example, the constructor of the \code{PBEKeySpec} class asks for an \code{iterationCount} parameter that specifies how many times the password is hashed to derive a crypto key~\cite{turan2010recommendation}.
Unfortunately the documentation does not indicate the minimum recommended value for this parameter.

Non-official documentation also suffers from similar issues.
There are many cryptography questions on the Stack Overflow website, but the answers may be out of date, or the proposed solutions may not consider security implications.
For instance, only fewer than 20\% of examined posts suggest a secure solution to implement SSL certificate checks, and a considerable number of developers did not comprehend the security concept of coding options, \eg disabling the CSRF token \cite{meng2018secure}.

Researchers actively attempt to improve the current state of using crypto APIs.
For instance,
a study of crypto uses in open-source projects claims that none of the factors such as the number or frequency of crypto-related code commits, or the number of projects that developers are involved in, correlate with developer performance in this domain~\cite{hazhirpasand2019impact}.
An experiment with 53 developers shows that API-integrated security hints help 73\% of developers to write more secure code~\cite{gorski2018developers}.
In the same vein, researchers have provided developers with an interactive web platform to access correct uses of crypto APIs~\cite{hazhirpasand2020cryptoexplorer}.
There also exist several static analysis tools such as CryptoLint~\cite{Egele}, and \CG~\cite{KrugerS0BM18} to help developers spot crypto API misuses in programs.

In this work, we conduct an exploratory study to understand the current state of crypto API usage in open-source Java projects.
In contrast to previous work, we investigate this topic at the API level \ie we explain which APIs are problematic and what types of misuses prevail.
We also contacted developers of these APIs to uncover the reasons underlying these misuses.
In particular,
we used the \CG static analysis tool to analyze hundreds of Java projects that rely on JCA APIs, and manually checked almost half of the results to understand the extent to which the tool result is reliable.
We observed that crypto misuses are common; particularly, the majority of APIs (\ie 13 of 15) were misused at least once.
For instance, \MD, the most prevalent JCA API in the projects, was misused in 92\% of the cases.
Our manual investigation revealed that the tool result is highly (\ie 93\%) reliable, however, in effect, the implications of a misuse depend on factors such as where and how the application is intended to be used.
For instance, a crypto misuse that makes software vulnerable when running on a network may be used only locally.
We prepared a publicly available dataset, called \CM, containing the details of the analysis, \eg project name, file name, API name, and line number.\footnote{\url{http://crypto-explorer.com/cryptomine}}
We created issues on the \GH repository of 216 projects to inform developers about crypto misuses.
We collected feedback concerning 140 issue reports, and two authors of this paper reviewed and classified them into nine major categories, \eg uncertain developers, non-security context, or lack of developer knowledge.
Analyzing the developer responses revealed that in many cases developers are well-informed about the right usage of crypto APIs, but the specific code context may force them to misuse certain crypto APIs.
However, other developers may be uncertain about correct API usage since security caveats in the crypto API documentation are scarce, or they used a third-party code that contained a misuse.


In the remainder of this paper, we describe the methodology of our work in \autoref{sec:method}.
Then, we describe the current status of the dataset and responses of developers in \autoref{sec:results}.
We point out potential threats to the validity of this work in \autoref{sec:threats}, and discuss related work in \autoref{sec:relwork}.
We conclude the paper in \autoref{sec:conclusion}.




\section{Methodology}
\label{sec:method}

We followed a few main steps
to analyze Java projects, build the \CM dataset, and collect developer feedback on the crypto misuses.
We explain these steps in the following.


\subsection{Searching, downloading, and compiling}
We start with a set of projects that use JCA APIs and were identified in previous work \cite{kruger2019crysl}.
We use \GH APIs to fetch the collaborators of these projects and check
what other Java projects they contributed to, which helps us to collect more projects.
Next, we use the \GH search code API to check whether a Java project uses any of the JCA APIs, such as \MD, \code{Cipher}, or \code{KeyStore}.
The \GH search code API limits the number of requests to 30 per minute, therefore we execute this phase in parallel with different \GH accounts.
If the project is forked, then we clone the original repository.
Forked projects significantly increase the chance of having duplicate projects in the dataset.
We also did not limit our search criteria based on the number of project forks or stars, as we were only interested in collecting crypto API uses regardless of factors such as project size, popularity, or the degree of recent activity.

We compile each project in preparation for the static analysis phase.
We use a bash script to check for the existence of the build file (POM) in the project's path and then proceed to compile the project using Maven.
We exclude any projects that cannot be compiled due to unresolved dependencies.

\subsection{Analysis}
We use an open-source static analysis tool called \CG, which detects known misuses of JCA APIs~\cite{KrugerS0BM18}.
It uses a set of rules to analyze method-call patterns, parameter constraints, and secure compositions of cryptography-related classes.
We chose \CG as it supports a wide range of APIs, is open-source, and relies on an extensive rule set created by crypto experts.
We extended the tool to collect and report information regarding at what line number each API is used and in which user-defined method the API use occurred.

We feed each project's binary code (\ie \textit{.class} files) to \CG.
Most projects are analyzed within 10 minutes.
Accordingly, we abort lengthy analyses that take more than 15 minutes.
Ultimately, 48 analyses were terminated.

For every successful analysis, we use the \GH API to obtain general information regarding each project, \ie the number of stars, the number of forks, the creation and the last updated date of the project.

\subsubsection{Schema}
We use a bash script to extract information from the generated analysis reports with the help of regular expressions.
We present the extracted values in the \CM dataset as a comma-separated CSV file.
Each record describes a single crypto API use.
\autoref{tbl:schema} presents each field and its description in a record.
Each data record represents meta-information about a crypto API use in a project such as line number of the API use, Java file path containing the API use, or project's address on \GH.
%

\begin{table}[h]
\small
\caption {Fields of each API use in CryptoMine} \label{tbl:schema}
\begin{tabular}{|l|p{6cm}|}
\hline
\multicolumn{1}{|c|}{\textbf{Field}} & \multicolumn{1}{c|}{\textbf{Description}}                                                                                                                                        \\ \hline
ps\_url                         &  project's address on GitHub                                                                                                                                                   \\ \hline
Star\_count                          & number of stars of a project                                                                                                                                                     \\ \hline
Fork\_count                         & number of forks of a project                                                                                                                                                     \\ \hline
Creation\_date                    & creation date of a project                                                                                                                                                       \\ \hline
Updated\_date                    & last updated date of a project                                                                                                                                                   \\ \hline
Last\_visited                       & the last time we checked a project                                                                                                                                                  \\ \hline
File\_path                           & file path containing a crypto use                                                                                                                                                \\ \hline
S\_object                            & status of a use (0 means a misuse, otherwise it is 1)                                                                                                                            \\ \hline
API\_name                  & name of the crypto API                                                                                                                                                           \\ \hline
Line\_number                         & line number of the crypto use                                                                                                                                                    \\ \hline
User\_method                         & \begin{tabular}[c]{@{}l@{}}the user-defined method where the crypto\\ API use is\end{tabular}                                                                                    \\ \hline
Misuse\_type                         & \begin{tabular}[c]{@{}l@{}}A string referring to the type of the \\ crypto misuse (wrong type, wrong object, \\wrong constraint,   incomplete operation, \\ incomplete order)
\end{tabular}
\\ \hline
Misuse\_desc                         & the description of the crypto misuse                                                                                                                                             \\ \hline
Manual\_check                        & \begin{tabular}[c]{@{}l@{}} the manually checked status of an API use \\ (Accepted, Rejected, Unvalidated)\end{tabular}                 \\ \hline
\end{tabular}
\end{table}

The user\_method field provides information about the user-defined function where a crypto API use exists.

The misuse\_type field can represent any of the following five types provided by the static analysis tool.
The ``wrong type'' means when a developer incorrectly uses a certain reference type.
For instance, the constructor of PBEKeySpec requires the password to be passed as a character array, and should not be as a string object.
The ``wrong object''  occurs when an object is passed to another object but not in the correct way to fulfill expected security requirements.
The ``wrong constraint'', which is a common misuse type, occurs when a developer selects wrong values for integer or string objects to pass to a crypto API, like key sizes, algorithm names, or iteration counts.
The ``incomplete operation'' indicates  the whole path for the desired cryptographic purpose is not fulfilled, \eg failing to call PBEKeySpec.clearPassword().
Finally, the ``incomplete order'' shows that the expected method call sequence to be made is incorrect, \eg failing to call to init() in the \code{Cipher} API.


The misuse\_desc field explains for what reason, which is provided by the tool,  a crypto API use violates \CG's rules.
The ``manual\_check'' field indicates the manual cross-validation status of an API use.
In case of approval, \ie agreement with the tool, we set the value of the field to \textit{Accepted}, otherwise, the value is set to \textit{Rejected}.
Non-validated records are indicated by \textit{Unvalidated}.
Lastly, interested researchers can request to receive the cloned version of the projects.

\subsubsection{Manual investigation}
Two authors of this paper manually checked 1280 records of \CM (48\% of the dataset).
They relied on their expertise and the CrySL rules provided by the static analysis tool.
The CrySL rules determine the secure uses of a crypto API.
The reviewers examine the 1280 records separately and finally cross-check their individual judgments.
In case of conflicts, they refer to the tool's rules and discuss them.

\subsection{Contacting the developers}
To understand the reasons behind the misuses, we contacted 216 maintainers of repositories on GitHub, which represents a sample size with a 95\% confidence level and 5\% margin of error.
For each repository, we opened an issue on the GitHub page, explained our objectives for reporting crypto misuses, provided an explanation for each misuse, and pinpointed the affected Java files, associated line numbers and API names.
We waited 20 days for responses from the developers of repositories, and then we manually extracted their responses.
Thereafter, two authors of this paper reviewed each response to determine the key message of each response.
Finally, they cross-checked their findings and, in case of a conflict, they revisited the concerned response.


\section{Results}
\label{sec:results}

In this section we first report on the current status of crypto API uses in open-source projects, and then present the key messages of developer feedback.

\subsection{The state of crypto uses}


We investigated the use of 15 JCA APIs in 489 projects.
We found that only two projects are completely healthy, and a staggering 487 projects suffer from at least one crypto misuse.
The mean of the project forks is 139, and the median value is 7.5.
The mean of the project stars is 348, and the median value is 5.

Among the manually investigated records, 74 records (6\%) were flagged as rejected, which means according to the tool's rules and the opinions of reviewers they are mistakenly marked as misuses.
For instance, before using the \code{sign} method in the \code{Signature} API, a developer needs to call either the \code{initSign} or the \code{update} method.
However, in some cases, developers used the \code{update} method in a loop, while the automatic analysis could not recognize it.

\begin{figure}
\centering
\includegraphics[width=1\linewidth,trim=4 4 4 4,clip]{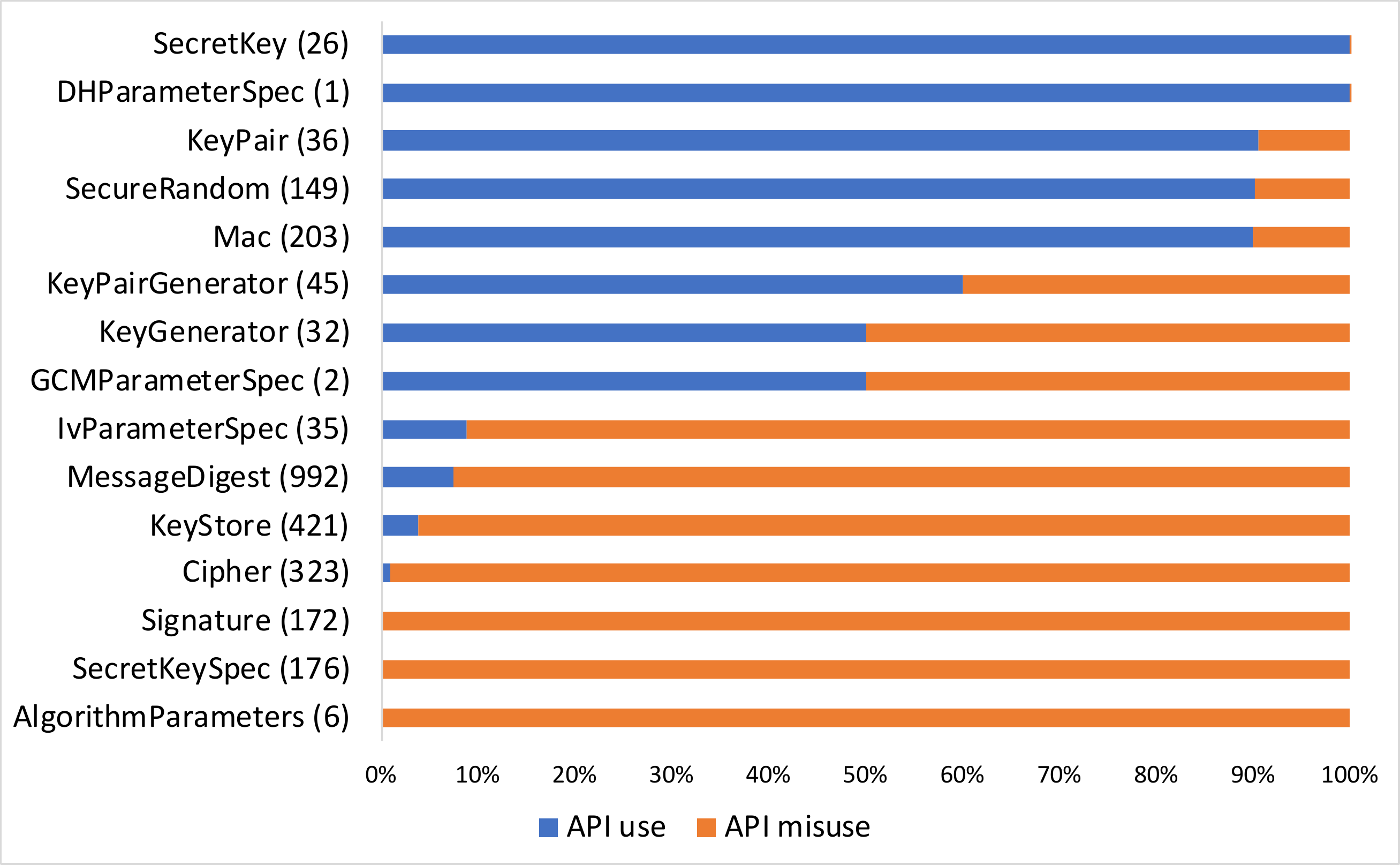}
\caption{The misuses \vs uses of each API in percentage}
\label{fig:chartjcaapi}
\end{figure}

Figure \ref{fig:chartjcaapi} summarizes the uses vs.\ misuses of each of these APIs as well as the total number of each API use in parenthesis.
Developers seemingly have severe difficulties in using more than half of the APIs whose correct uses were less than 51\%.
For instance, The correct uses of five APIs namely, \code{SecretKeySpec}, \code{Iv\-Para\-meter\-Spec}, \code{KeyStore}, \code{Cipher}, \MD, and \code{Signature} were at most 10\%.
In contrast, developers had a promising performance in using the \code{SecretKey}, \code{Mac}, \code{SecureRandom}  and \code{KeyPair} APIs, \ie at least 90\% uses were correct.




Various misuse types may pose threat with different levels of severity depending on a project is intended to be used, \eg selecting a wrong constraint MD5 versus skipping to dispose of a crypto object.
\autoref{tab:meta} gives information about the distribution of crypto misuse types in the top six most misused APIs.
Most of the misuses were of the \code{ConstraintError} type followed by \code{RequiredPredicateError} and \code{Typestate\-Error}.
%
The \code{ConstraintError} type made up the largest proportion of misuses, showing that developers struggle with choosing correct parameters for crypto APIs.
The second most common misuse is the \code{RequiredPredicateError} type, which means an insecure object is passed to other objects as an arguments.
The \code{NeverTypeOfError} and \code{IncompleteOperationError} types account for nearly 23\% of the total misuse types that exist in all analyzed projects.

The complete analysis results are publicly available via the \CM dataset,\footnote{\url{http://crypto-explorer.com/cryptomine}} which facilitates investigation of the following research questions:
(1) What are the most common crypto mistakes, and what should we do to improve learnability in this domain?
(2) How do crypto uses evolve in a project?
(3) How do the quality characteristics of a project correlate with crypto uses in that project?
(4) In what context are crypto APIs commonly used?
(5) Why do some developers perform better in using cryptography?
(6) What is the performance of the static analysis tool in detecting crypto misuses?
(7) What is the benchmark result of comparing several static analysis tools in detecting crypto uses of the dataset's projects?

\begin{table*}[ht]
\centering
\caption {Mostly misused APIs with more than 10 misuse types} \label{tab:meta}
\begin{adjustbox}{max width=\textwidth} 
\small
\begin{tabular}{llllll} \hline
\textbf{JCA API}                           & \textbf{IncompleteOperationError} & \textbf{NeverTypeOfError} & \textbf{TypestateError} & \textbf{RequiredPredicateError} & \textbf{ConstraintError} \\ \hline
SecretKeySpec      &                                   &                           &                         & 170                             &                          \\
Signature          & 5                                 &                           & 1                       & 49                              & 26                       \\
Cipher              & 42                                &                           &                         & 83                              & 52                       \\
KeyStore         & 32                                & 160                       & 16                      &                                 &                          \\
MessageDigest      & 72                                &                           & 178                     &                                 & 400                      \\
IvParameterSpec   &                                   &                           &                         & 30                              &                          \\
\hline
\end{tabular}
\end{adjustbox}
\end{table*}
\subsection{Developer feedback}
\label{sec:misuse}
Out of 216 repository maintainers, 76 did not respond, and 140 reacted to the issues within 20 days.
Among all the reported misused APIs,
\MD had the greatest number of submitted and received responses.
\code{KeyStore} and \code{SecretKeySpec} are respectively in the second and third place with 22 and 15 received responses from 32 and 31 submitted issues.
Following that, \code{Cipher} and \code{Signature} were the last two APIs that had been reported in more than 10 submissions.
The rest of the APIs had only a few submission and received responses.



We evaluated all of the responses for 140 repositories in order to identify developer perceptions concerning cryptographic APIs.
This widens our views on how knowledgeable developers are when they misuse a crypto API in their code.

We realized that only a tiny fraction of repository maintainers (\ie of seven repositories) agreed to fix the issues, and a large number of maintainers (\ie of 46 repositories) disagreed since the context where the misuse occurred was not considered to be security-sensitive.
Fortunately, 32  repository maintainers were interested in starting a dialogue about exactly why a given issue can cause problems, and whether the associated risks can arise in practice.
We present nine main categories of responses in the following and highlight the key findings of each category in a box.


\paragraph{\textbf{Personal repository}}
We received three responses indicating that the target repository is for a personal use.
A contributor said that \textit{``the project is meant to be used for educational purposes and intentionally contains some vulnerable examples.''}
Another mentioned that \textit{``the project is created for internal use and no issue will be addressed.''}

\boxit{
People are not aware of the impact these issues could have on those who rely on online examples, as their repositories are publicly accessible.
Another facet could be that they are not concerned about security when a program is being used on a very small scale.
}

\paragraph{\textbf{Will fix later}}
Developers of seven repositories replied that they will fix the reported crypto misuses later without asking for any further explanation.
Three of them replied that those misuses do not affect the functionality of the program and are not urgent to be fixed.

\boxit{
Developers often underestimate the impact of a crypto misuse.
}

\paragraph{\textbf{Request for pull}}
Developers of 17 repositories suggested to create a pull request.
For instance, a contributor responded that \textit{``I'm not sure if I understand the problem.
I am not a cryptologist.''}
We believe a lack of knowledge in this area exists that may cause developers to blindly accept a pull request.
The inevitable consequence of blindly accepting a pull request could adversely affect the security of the final software, for instance, an adversary may submit a downgrade to the existing security mechanisms in a project.

\boxit{
There is a risk that developers who lack security knowledge blindly accept security-related pull requests.
}

\paragraph{\textbf{Refer to the main library}}
In five cases developers used an open-source library, and asked us to report the issue to the library's repository.

\boxit{
Unfortunately, developers seem not to be concerned about security risks associated with external libraries.
}

\paragraph{\textbf{Repo is not maintained}}
We learned from the responses that 15 repositories are not maintained anymore.

\boxit{Inactive projects are common in the open source community, for example due to a lack of financial support.
However, as long as the code is available online, novice developers may rely on open-source projects irrespective of how active the projects are.}

\paragraph{\textbf{Consult documentation}}
Developers of 10 repositories were not completely certain about how the APIs should be used securely.
They either asked us to read the API documentation or quoted a relevant part of the documentation in their responses.
For instance, we suggested not to use java.lang.String as the second parameter of \code{KeyStore}, which is the password parameter.
They replied that \textit{``according to the documentation, the parameter is a String, so why should it never be?''}
One responded that  \textit{``MD5 is still supported by java according to the Java documentation.''}
Another one asked us to provide him with the correct use of \code{Signature} API as he did not know how to fix it.
%
A developer referred to the official Java documentation to express his trust in using the \textit{SHA1withRSA} algorithm in the \code{Signature} API.
Developers of two repositories were not convinced to stop using \code{NoPadding} in the \code{Cipher} API.
They noted that using an \textit{empty string} in Java 8 can cause a run-time error and they cited \code{Cipher}'s page in Java documentation.

\boxit{
Developers have confidence in official documentation, but security concerns are mainly absent in such resources.
}


\paragraph{\textbf{Uncertainty}}
We found that many developers (\ie of 32 repositories) asked us to provide a clarification.
Some developers referred to blog posts where the \code{KeyStore} API was misused by converting a string variable to an array of characters, \ie \textit{password.toCharArray()}, and passing it as the second parameter to the API.
A common skepticism was about which algorithm is safe to use in SecretKeySpec and KeyPairGenerator.

A few developers asked how the misuses can be exploited in real life.
For instance, a contributor was not convinced about the nature of the ``misuse'' as it was not clear to him how a wrong transformation mode in the \code{Cipher} API can be exploited in his application.

\boxit{As expected, developer uncertainty regarding the correct way of using an API securely is related to either the right method call or the secure algorithm name}


\paragraph{\textbf{Consider the context and disagreement}}

The majority of responses (46) are connected with the context of the code.
A large number of these responses were mainly related to the \MD API since \MD was used more frequently than any other crypto API in the analyzed projects.
This is because \MD can be used in many different scenarios such as authentication, checksumming, archiving, or in combination with other algorithms.

One common complaint was that MD5 or SHA1 were not being used for security purposes.
They had been used for archiving or producing hashes for non-security use-cases.
For instance, one developer mentioned that the Redis API needs to generate checksums using SHA1.
As another example of non-security usage, one repository used SHA1 for opening a handshake in WebSocket and the contributor referred to the RFC 6455 section 1.3 for further information.
Moreover, three repository maintainers replied that instead of using \code{String.hashCode()}, they used SHA1/MD5 as an internal identifier, \ie generating a normalized document ID based on the URL of the given document.
Another developer stressed that they use MD5 in order to track if the template source has been changed or not.
A group of developers used MD5 to get a hash of an email address to produce the avatar URL of the user.
SHA1 was also used in a \code{for} loop to generate fake data to be stored in a file in a repository.
Contributors to a repository  pointed to a code comment preceding the SHA1 usage that clearly says that \textit{SHA1 was used only to generate a single hash for the entire contents of a folder and it is absolutely sufficient}.

Some contributors complained that the critiqued code is very old and the context is not security-focused, \ie a decade old, and running a static analysis is not a good measure to find misuses.
In some responses, they did not exactly mention what the context was and only replied that the context is not security-sensitive.
For example, we manually checked the codes of the repositories and found that \MD was used for purposes such as hashing parameters, \ie album name, in the URL or to cache the unpacked ZIP file and avoid multiple extractions.

In one repository, developers indicated that there was a code comment explaining why the \code{Signature} API had been used in an insecure way.
A developer mentioned that although \code{KeyPairGenerator} accepts many algorithms such as RSA, DSA, and Diffie-Hellman, in this project the APNs protocol demands \code{KeyPairGenerator} to generate key pairs for the Elliptic Curve algorithm.

One contributor cited a blog post where the blogger discussed that SHA1 is still usable regardless of the existing collision vulnerability.
He insisted that they will continue using the algorithm until a serious security problem is raised by using SHA1.
Another developer stressed that MySQL authentication plugins do not support the usage of SHA-256 and accordingly, SHA1 was used in their project.
In contrast, official MySQL documentation added that since MySQL 5.6, the sha256\_password authentication plugin is supported.
With regard to the misuse of MD5, a developer replied that he cannot change the algorithm name as the remote endpoint requires an MD5 hash and he is not able to change it on the remote endpoint.

\boxit{
Developers mainly argued that their context is not related to security.
The use of security APIs to produce hashes was the most common non-security related usage.
}
\\

We witnessed that some APIs were more prevalent compared to others in our reports.
For example, the \MD API was seen more than other APIs in all response categories except for ``Personal repository.''
In \MD, the most common misuse type is \emph{constraint error} in which developers used the MD5 or SHA1 algorithms to compute hashes.
As hashing algorithms could be used for non-security purposes, many maintainers were expected to note that the context is not relevant.
\code{KeyStore} and \code{Cipher} are the second most frequently seen APIs in the responses.
Repository maintainers mostly asked for clarification or referred to the documentation of Java for the misuses of \code{KeyStore}.
The \code{Cipher} API had the majority of misuses linked to its first argument.
This occurred due to the diversity of options in the transformation string.
A transformation string includes the name of a cryptographic algorithm (e.g., AES or DES), and may be followed by a feedback mode and padding scheme, \eg "algorithm/mode/padding".
On the whole, developers had difficulty in understanding what constraint they should pass to the crypto APIs or how to create an object securely in order to pass it to another crypto API.

The responses of maintainers highlight the fact that some developers are fully aware of what they are doing, whereas others have doubts concerning the correct way of using such APIs.
As a result, blaming only developers for the found crypto API misuses cannot be correct.
To highlight the leading causes of crypto misuses, we identify various influential factors that must be taken into account.
The \emph{age of a project} can be an essential factor as security standards can evolve over time.
Another factor is to blindly rely on the \emph{use of third-party libraries}.
Developers need to spend more time and select libraries with higher credibility and support.
\emph{Official documentation} and
\emph{unofficial online documentation}, such as Stack Overflow, can have pernicious effects on developer choice.
This impact can be ruinous when the developer lacks the minimum knowledge in the domain of cryptography, \eg choosing ECB mode from the examples provided by the official documentation.
Furthermore, such developers may not be able to make a good use of static analysis tools to resolve their crypto problems.
On the other hand, \emph{statically checking the developer's code without considering the context} yields misleading results and could not be a dependable measure for developer performance in this domain.
For instance, recent program analysis tools consider SHA-1 to be insecure.
Such tools produce alerts if a developer cautiously uses three nested SHA-1 to produce a hash while the application is not, indeed, exposed to security threats.
Lastly, even though some developers were aware of the right crypto usage,
the widespread of such examples on open-source projects may have profound implications on inexperience developers.

\section{Threats to Validity}
\label{sec:threats}
It is infeasible to manually identify crypto (mis)uses in source code at large-scale.
Adoption of static code analysis tools can considerably help developers to automatically detect crypto misuses and write more secure code.
Therefore, we employed a static analysis tool, \ie \CG, in order to assess the status of crypto uses in hundreds of Java projects.
The primary reason for this choice is that the tool is open-source and supports a wide range of crypto rules for different APIs, which are easily extendable.
However, the \CM dataset does not represent all the JCA APIs and their various usages, \eg different parameters or method calls.
This can be addressed by increasing the number of analyzed projects which contain different usages of crypto APIs.
To provide a better level of reliability, we have manually cross-checked 48\% of the results in the \CM dataset.
Nevertheless, the manual analysis of a large dataset is a non-trivial task and to accelerate the process we invite interested researchers to join us.
Moreover, security assumptions may change over a period of time.
For instance SHA1, which was judged to be secure in the past, is considered insecure now.
Further work is needed to add an extensive range of API usages and to contact more repositories to collect more evidence for the corresponding API misuses.
\section{Related Work}
\label{sec:relwork}
A series of recent studies have indicated that developers need more security support from different aspects, \eg secure documentation, secure sample code, to write secure code.
Nadi \etal surveyed two groups of developers, \ie  11 developers who had crypto-related questions on Stack Overflow, and 37 developers who used Java's cryptography APIs.
The authors realized that developers are certain in choosing the proper cryptography concepts, but problems still exist in using certain cryptographic algorithms correctly.
They arrived at the conclusion that crypto APIs are generally regarded to be too low-level, and developers choose more task-based solutions~\cite{nadi2016jumping}.
In a study, Acar \etal asked 54 participants to solve pre-defined challenges in which the participants had to write security- and privacy-relevant code under time limitations \cite{acar2016you}.
The authors noticed that many of the security problems made by
their participants also can be found on online sources, \eg Stack Overflow.
Conducting a controlled experiment with 53 participants,
Gorski \etal showed the effectiveness of API-integrated security advice can significantly help (\ie 73\%) to produce more secure code \cite{gorski2018developers}.

Previous studies have emphasized that developers commonly misused crypto APIs.
Rahman \etal used their static analysis tool, named CHIRON, 
to evaluate 46 large-scale Apache projects
\cite{Rahaman2018chiron}.
They found a total of 2,009 alerts in the projects.
Egele \etal developed a static analysis tool, \ie CryptoLint, and tested on 11,748 applications using cryptographic APIs~\cite{Egele}.
Their results indicated that 88\% of the applications used crypto APIs inappropriately.
Kr\"uger \etal introduced a tool called \CG, an Eclipse plugin that assists developers to securely use cryptographic API \cite{kruger2019crysl}.
\CG also presents secure code templates so as to reduce the hassle of searching for secure API usages.
Besides, \CG employs CrySL that empowers cryptographic specialists to expand the rules for other APIs\cite{KrugerS0BM18}.
Gao \etal made an assumption that developers update API usage instances to fix misuses, and accordingly, conducted a large scale analysis on nearly 40 000 real-world app lineages to trace API usage rules \cite{gao2019negative}. They failed to confirm their assumption that API usage updates tend to fix misuses.



\section{Conclusion}
\label{sec:conclusion}
We analyzed hundreds of projects in which JCA APIs were used, to observe the status of API use in open-source projects, to learn what crypto misuse types exist, and to investigate the influential factors in misusing such APIs.
We found that 85\% of the crypto APIs suffered from at least one misuse, though not all misuses were at the same level of severity.
We contacted the maintainers of the projects to understand the reasons behind the misuse of crypto APIs, and we classified their responses into nine main categories.
The results demonstrate that security hints in API documentation are scarce, misuses are rooted in third-party libraries, or the code context plays a crucial role in using crypto APIs incorrectly.
Finally, to support the research community, we publicly share the \CM dataset, including the analysis results, and information about each project such as its metadata information, the precise locations of API use, and the safety status of these APIs, to name but a few.


\section*{Acknowledgment}
We gratefully acknowledge the financial support of the
Swiss National Science Foundation for the project ``Agile Software Assistance'' (SNSF project No.\,200020-181973, Feb.\,1, 2019 -- April 30, 2022).


\bibliographystyle{ACM-Reference-Format}
\bibliography{thebibliography}


\begin{thebibliography}{12}


\ifx \showCODEN    \undefined \def \showCODEN     #1{\unskip}     \fi
\ifx \showDOI      \undefined \def \showDOI       #1{#1}\fi
\ifx \showISBNx    \undefined \def \showISBNx     #1{\unskip}     \fi
\ifx \showISBNxiii \undefined \def \showISBNxiii  #1{\unskip}     \fi
\ifx \showISSN     \undefined \def \showISSN      #1{\unskip}     \fi
\ifx \showLCCN     \undefined \def \showLCCN      #1{\unskip}     \fi
\ifx \shownote     \undefined \def \shownote      #1{#1}          \fi
\ifx \showarticletitle \undefined \def \showarticletitle #1{#1}   \fi
\ifx \showURL      \undefined \def \showURL       {\relax}        \fi
\providecommand\bibfield[2]{#2}
\providecommand\bibinfo[2]{#2}
\providecommand\natexlab[1]{#1}
\providecommand\showeprint[2][]{arXiv:#2}

\bibitem[\protect\citeauthoryear{Acar, Backes, Fahl, Kim, Mazurek, and
  Stransky}{Acar et~al\mbox{.}}{2016}]%
        {acar2016you}
\bibfield{author}{\bibinfo{person}{Yasemin Acar}, \bibinfo{person}{Michael
  Backes}, \bibinfo{person}{Sascha Fahl}, \bibinfo{person}{Doowon Kim},
  \bibinfo{person}{Michelle~L Mazurek}, {and} \bibinfo{person}{Christian
  Stransky}.} \bibinfo{year}{2016}\natexlab{}.
\newblock \showarticletitle{You get where you're looking for: The impact of
  information sources on code security}. In \bibinfo{booktitle}{\emph{2016 IEEE
  Symposium on Security and Privacy (SP)}}. IEEE, \bibinfo{pages}{289--305}.
\newblock


\bibitem[\protect\citeauthoryear{Egele, Brumley, Fratantonio, and
  Kruegel}{Egele et~al\mbox{.}}{2013}]%
        {Egele}
\bibfield{author}{\bibinfo{person}{Manuel Egele}, \bibinfo{person}{David
  Brumley}, \bibinfo{person}{Yanick Fratantonio}, {and}
  \bibinfo{person}{Christopher Kruegel}.} \bibinfo{year}{2013}\natexlab{}.
\newblock \showarticletitle{An Empirical Study of Cryptographic Misuse in
  Android Applications}. In \bibinfo{booktitle}{\emph{Proceedings of the 2013
  ACM SIGSAC Conference on Computer \& Communications Security}} (Berlin,
  Germany) \emph{(\bibinfo{series}{CCS '13})}. \bibinfo{publisher}{ACM},
  \bibinfo{address}{New York, NY, USA}, \bibinfo{pages}{73--84}.
\newblock
\showISBNx{978-1-4503-2477-9}
\urldef\tempurl%
\url{https://doi.org/10.1145/2508859.2516693}
\showDOI{\tempurl}


\bibitem[\protect\citeauthoryear{Gao, Kong, Li, Bissyand{\'e}, and Klein}{Gao
  et~al\mbox{.}}{2019}]%
        {gao2019negative}
\bibfield{author}{\bibinfo{person}{Jun Gao}, \bibinfo{person}{Pingfan Kong},
  \bibinfo{person}{Li Li}, \bibinfo{person}{Tegawend{\'e}~F Bissyand{\'e}},
  {and} \bibinfo{person}{Jacques Klein}.} \bibinfo{year}{2019}\natexlab{}.
\newblock \showarticletitle{Negative results on mining crypto-API usage rules
  in Android apps}. In \bibinfo{booktitle}{\emph{Proceedings of the 16th
  International Conference on Mining Software Repositories}}. IEEE Press,
  \bibinfo{pages}{388--398}.
\newblock


\bibitem[\protect\citeauthoryear{Gorski, Iacono, Wermke, Stransky, M{\"o}ller,
  Acar, and Fahl}{Gorski et~al\mbox{.}}{2018}]%
        {gorski2018developers}
\bibfield{author}{\bibinfo{person}{Peter~Leo Gorski}, \bibinfo{person}{Luigi~Lo
  Iacono}, \bibinfo{person}{Dominik Wermke}, \bibinfo{person}{Christian
  Stransky}, \bibinfo{person}{Sebastian M{\"o}ller}, \bibinfo{person}{Yasemin
  Acar}, {and} \bibinfo{person}{Sascha Fahl}.} \bibinfo{year}{2018}\natexlab{}.
\newblock \showarticletitle{Developers Deserve Security Warnings, Too: On the
  Effect of Integrated Security Advice on Cryptographic API Misuse}. In
  \bibinfo{booktitle}{\emph{Fourteenth Symposium on Usable Privacy and Security
  SOUPS 2018)}}. \bibinfo{pages}{265--281}.
\newblock


\bibitem[\protect\citeauthoryear{Hazhirpasand, Ghafari, Kr{\"u}ger, Bodden, and
  Nierstrasz}{Hazhirpasand et~al\mbox{.}}{2019}]%
        {hazhirpasand2019impact}
\bibfield{author}{\bibinfo{person}{Mohammadreza Hazhirpasand},
  \bibinfo{person}{Mohammad Ghafari}, \bibinfo{person}{Stefan Kr{\"u}ger},
  \bibinfo{person}{Eric Bodden}, {and} \bibinfo{person}{Oscar Nierstrasz}.}
  \bibinfo{year}{2019}\natexlab{}.
\newblock \showarticletitle{The Impact of Developer Experience in Using Java
  Cryptography}. In \bibinfo{booktitle}{\emph{the International Symposium on
  Empirical Software Engineering and Measurement, {ESEM} 2019}}.
\newblock


\bibitem[\protect\citeauthoryear{Hazhirpasand, Ghafari, and
  Nierstrasz}{Hazhirpasand et~al\mbox{.}}{2020}]%
        {hazhirpasand2020cryptoexplorer}
\bibfield{author}{\bibinfo{person}{Mohammadreza Hazhirpasand},
  \bibinfo{person}{Mohammad Ghafari}, {and} \bibinfo{person}{Oscar
  Nierstrasz}.} \bibinfo{year}{2020}\natexlab{}.
\newblock \showarticletitle{CryptoExplorer: An Interactive Web Platform
  Supporting Secure Use of Cryptography APIs}. In
  \bibinfo{booktitle}{\emph{27th IEEE International Conference on Software
  Analysis, Evolution and Reengineering {SANER} 2020}}.
\newblock


\bibitem[\protect\citeauthoryear{Kr{\"{u}}ger, Sp{\"{a}}th, Ali, Bodden, and
  Mezini}{Kr{\"{u}}ger et~al\mbox{.}}{2018}]%
        {KrugerS0BM18}
\bibfield{author}{\bibinfo{person}{Stefan Kr{\"{u}}ger},
  \bibinfo{person}{Johannes Sp{\"{a}}th}, \bibinfo{person}{Karim Ali},
  \bibinfo{person}{Eric Bodden}, {and} \bibinfo{person}{Mira Mezini}.}
  \bibinfo{year}{2018}\natexlab{}.
\newblock \showarticletitle{CrySL: An Extensible Approach to Validating the
  Correct Usage of Cryptographic APIs}. In \bibinfo{booktitle}{\emph{32nd
  European Conference on Object-Oriented Programming, {ECOOP} 2018, July 16-21,
  2018, Amsterdam, The Netherlands}}. \bibinfo{pages}{10:1--10:27}.
\newblock


\bibitem[\protect\citeauthoryear{Kr{\"u}ger, Sp{\"a}th, Ali, Bodden, and
  Mezini}{Kr{\"u}ger et~al\mbox{.}}{2019}]%
        {kruger2019crysl}
\bibfield{author}{\bibinfo{person}{Stefan Kr{\"u}ger},
  \bibinfo{person}{Johannes Sp{\"a}th}, \bibinfo{person}{Karim Ali},
  \bibinfo{person}{Eric Bodden}, {and} \bibinfo{person}{Mira Mezini}.}
  \bibinfo{year}{2019}\natexlab{}.
\newblock \showarticletitle{Crysl: An extensible approach to validating the
  correct usage of cryptographic apis}.
\newblock \bibinfo{journal}{\emph{IEEE Transactions on Software Engineering}}
  (\bibinfo{year}{2019}).
\newblock


\bibitem[\protect\citeauthoryear{Meng, Nagy, Yao, Zhuang, and Argoty}{Meng
  et~al\mbox{.}}{2018}]%
        {meng2018secure}
\bibfield{author}{\bibinfo{person}{Na Meng}, \bibinfo{person}{Stefan Nagy},
  \bibinfo{person}{Danfeng Yao}, \bibinfo{person}{Wenjie Zhuang}, {and}
  \bibinfo{person}{Gustavo~Arango Argoty}.} \bibinfo{year}{2018}\natexlab{}.
\newblock \showarticletitle{Secure coding practices in java: Challenges and
  vulnerabilities}. In \bibinfo{booktitle}{\emph{Proceedings of the 40th
  International Conference on Software Engineering}}.
  \bibinfo{pages}{372--383}.
\newblock


\bibitem[\protect\citeauthoryear{Nadi, Kr{\"u}ger, Mezini, and Bodden}{Nadi
  et~al\mbox{.}}{2016}]%
        {nadi2016jumping}
\bibfield{author}{\bibinfo{person}{Sarah Nadi}, \bibinfo{person}{Stefan
  Kr{\"u}ger}, \bibinfo{person}{Mira Mezini}, {and} \bibinfo{person}{Eric
  Bodden}.} \bibinfo{year}{2016}\natexlab{}.
\newblock \showarticletitle{Jumping through hoops: Why do Java developers
  struggle with cryptography APIs?}. In \bibinfo{booktitle}{\emph{Proceedings
  of the 38th International Conference on Software Engineering}}.
  \bibinfo{pages}{935--946}.
\newblock


\bibitem[\protect\citeauthoryear{Rahaman, Xiao, Tian, Shaon, Kantarcioglu, and
  Yao}{Rahaman et~al\mbox{.}}{2018}]%
        {Rahaman2018chiron}
\bibfield{author}{\bibinfo{person}{Sazzadur Rahaman}, \bibinfo{person}{Ya
  Xiao}, \bibinfo{person}{Ke Tian}, \bibinfo{person}{Fahad Shaon},
  \bibinfo{person}{Murat Kantarcioglu}, {and} \bibinfo{person}{Danfeng Yao}.}
  \bibinfo{year}{2018}\natexlab{}.
\newblock \showarticletitle{CHIRON: Deployment-quality Detection of Java
  Cryptographic Vulnerabilities}.
\newblock \bibinfo{journal}{\emph{arXiv preprint arXiv:1806.06881}}
  (\bibinfo{year}{2018}).
\newblock


\bibitem[\protect\citeauthoryear{Turan, Barker, Burr, and Chen}{Turan
  et~al\mbox{.}}{2010}]%
        {turan2010recommendation}
\bibfield{author}{\bibinfo{person}{Meltem~S{\"o}nmez Turan},
  \bibinfo{person}{Elaine Barker}, \bibinfo{person}{William Burr}, {and}
  \bibinfo{person}{Lily Chen}.} \bibinfo{year}{2010}\natexlab{}.
\newblock \showarticletitle{Recommendation for password-based key derivation}.
\newblock \bibinfo{journal}{\emph{NIST special publication}}
  \bibinfo{volume}{800} (\bibinfo{year}{2010}), \bibinfo{pages}{132}.
\newblock


\end{thebibliography}

\end{document}